\documentclass[showpacs,showkeys,11pt,
preprint,preprintnumbers,nofootinbib,
groupedaddress,superscriptaddress,amsmath,amssymb]{revtex4}

\usepackage[dvipdfm]{graphicx}
\usepackage{color}

\usepackage{ulem}

\newcommand{\red}[1]{{\color[rgb]{1,0,0} #1}}
\newcommand{\green}[1]{{\color[rgb]{0,0.6,0} #1}}
\newcommand{\blue}[1]{{\color[rgb]{0,0,1} #1}}

\newcommand{\rg}[1]{{\color[rgb]{0.7,0.7,0} #1}}
\newcommand{\rb}[1]{{\color[rgb]{1,0,1} #1}}
\newcommand{\gb}[1]{{\color[rgb]{0,0.8,0.8} #1}}

\newcommand{\SU}{{\text{SU}}}

\newcommand{\LQ}{{\text{LQ}}}
\newcommand{\DQ}{{\text{DQ}}}

\newcommand{\MNS}{{\text{MNS}}}
\newcommand{\CKM}{{\text{CKM}}}

\newcommand{\GeV}{{\text{GeV}}}
\newcommand{\TeV}{{\text{TeV}}}
\newcommand{\qd}{{i}}
\newcommand{\qdp}{{j}}

\begin{document}
\title{Lepton number violation at the LHC
with leptoquark and diquark}
\preprint{UT-HET 073}
\author{Masaya Kohda}
\email{mkohda@hep1.phys.ntu.edu.tw}
\affiliation{
Department of Physics, National Taiwan University,
Taipei 10617, Taiwan
}
\author{Hiroaki Sugiyama}
\email{sugiyama@sci.u-toyama.ac.jp}
\affiliation{
Department of Physics, University of Toyama,
Toyama 930-8555, Japan
} 
\author{Koji Tsumura}
\email{ko2@eken.phys.nagoya-u.ac.jp}
\affiliation{
Department of Physics, Graduate School of Science,
Nagoya University, Nagoya 464-8602, Japan}


\begin{abstract}

 We investigate a model in which
tiny neutrino masses are generated at the two-loop level
by using scalar leptoquark and diquark multiplets.
 The diquark can be singly produced at the LHC,
and it can decay into a pair of leptoquarks
through the lepton number violating interaction.
 Subsequent decays of the two leptoquarks can
provide a clear signature of the lepton number violation,
namely two QCD jets and a pair of same-signed charged leptons
without missing energy. 
 We show that the signal process is not suppressed
while neutrino masses are appropriately suppressed.
\end{abstract}
\maketitle

\section{Introduction} 


 The Standard Model~(SM) gauge symmetry of elementary particles 
based on the $\SU(3)_C \times \SU(2)_L \times \text{U}(1)_Y$
has been tested very accurately. 
 On the other hand,
the existence of the neutrino masses
has been established%
~\cite{Ref:solar-v,Ref:atom-v,Ref:acc-v,Ref:acc-app-v,
Ref:short-reac-v,Ref:long-reac-v}.
 This is clear evidence of the new physics beyond the SM
because neutrinos are massless in the SM\@.
 Since neutrinos are electrically neutral,
they can be Majorana particles
unlike the other SM fermions~\cite{Ref:Majorana}.
 The reason why neutrino masses are very different
from those of the other SM fermions
might be the Majorana property of neutrinos.

 The most familiar utilization of the Majorana property
to generate tiny neutrino masses
is the so-called Type-I Seesaw mechanism 
in which $\SU(2)_L$-singlet right-handed neutrinos
mediate in the tree diagram~\cite{Ref:Type-I}.
 Because of the suppression by mass scales of new heavy particles, 
naturally light neutrinos can arise.
 Another typical prescription
to obtain tiny Majorana neutrino masses
is the so-called radiative seesaw mechanism%
~\cite{Ref:Zee,Ref:Zee-Babu,Ref:Ma,Krauss:2002px,Aoki:2008av},
where neutrino masses are induced at the loop level.
 In these models,
the suppression of neutrino masses
can be achieved by the loop suppression factor
and/or a combination of new coupling constants
even if new particles are not very heavy.
 The masses of charged leptons
involved in the chirality flipping loop
provide further suppression of the neutrino masses
in some of such models%
~\cite{Ref:Zee,Ref:Zee-Babu,Krauss:2002px}.


 Although the lepton number is conserved in the SM,
the addition of the Majorana mass term of neutrinos
breaks the lepton number conservation by two units.
 The measurement of the lepton number violating~(L\#V) processes
such as the neutrinoless double beta decay%
~\cite{Ref:0v2beta,Ref:0v2beta-exp}
is extremely important
because it gives evidence that neutrinos are Majorana particles.
 Such processes are naively expected to be very rare
because neutrino masses are very small.
 This is true for the Type-I seesaw model
with very heavy right-handed neutrinos
because light Majorana neutrino masses are
unique lepton number breaking parameters
at the energy scale which is experimentally accessible.
 However,
in radiative seesaw models,
a trilinear coupling constant for light~(e.g.\ TeV-scale) scalars
can be more fundamental than light neutrino masses
as the L\#V parameter at the accessible energy scale.
 Then,
L\#V processes via the trilinear coupling constant
can be significant at the TeV-scale
even if the neutrino masses are suppressed enough.


 New particles related to the neutrino mass generation
are usually produced via the electroweak interaction,
and therefore the production cross sections are not so significant
at the LHC\@.
 However,
new particles in the loop of the radiative seesaw models
can be charged under the $\SU(3)_C$~\cite{Ref:cZM,Ref:cMM,Ref:cZBM}. 
 Such a colored particle can easily be produced at hadron colliders.
 In these models,
decay patterns of new colored particles
could be related to the form of the neutrino mass matrix
constrained by the neutrino oscillation data%
~\cite{Ref:cZM,Ref:cMM,Ref:cZBM,Ref:cMM-collider}
(See also \cite{Choubey:2012ux}).


 In this paper,
we investigate a radiative seesaw model
with a scalar leptoquark multiplet and a scalar diquark multiplet.
 Majorana masses of neutrinos are induced via the two-loop diagram
where colored particles are involved in the loop.
 The lepton number violation is caused
by the trilinear coupling constant of the leptoquarks and diquark, 
which can produce a characteristic signature at the LHC\@.
 The signature consists of
two QCD jets and a pair of same-signed charged leptons
without missing energy, 
which would be easily observed at the LHC\@.

 This paper is organized as follows. 
 In Sec.~II,
we present the model.
 Section~III
is devoted to discussion on the collider phenomenology
and the low energy constraints for the leptoquark and the diquark in the model.
 Conclusions are given in Section~IV\@.

\section{The Model}

\begin{table}[t]
\begin{center}
\begin{tabular}{c||c|c|c|c|c|c||c|c}
 {}
 &
  \ $L_\ell =
     \begin{pmatrix}
      \nu_{\ell L}^{}\\
      \ell_L
     \end{pmatrix}$ \
 &
  \ $Q_i^\alpha =
     \begin{pmatrix}
      u_{iL}^{\prime\,\alpha}\\
      d_{iL}^{\,\alpha}
     \end{pmatrix}$ \
 &
  \ $\Phi =
     \begin{pmatrix}
      \phi^+\\
      \phi^0
     \end{pmatrix}$ \
 & \ $\ell_R$ \
 & \ $u_{iR}^\alpha$ \
 & \ $d_{iR}^{\,\alpha}$ \
 & \ $S_\LQ^\alpha$ \
 & \ $S_\DQ^{\alpha\beta}$ \
\\\hline\hline
 \ $\text{SU}(3)_C$ \
 & \ {\bf \underline{1}} \
 & \ {\bf \underline{3}} \
 & \ {\bf \underline{1}} \
 & \ {\bf \underline{1}} \
 & \ {\bf \underline{3}} \
 & \ {\bf \underline{3}} \
 & \ {\bf \underline{3}} \
 & \ {\bf \underline{6}} \
\\\hline
 \ $\text{SU}(2)_L$ \
 & \ {\bf \underline{2}} \
 & \ {\bf \underline{2}} \
 & \ {\bf \underline{2}} \
 & \ {\bf \underline{1}} \
 & \ {\bf \underline{1}} \
 & \ {\bf \underline{1}} \
 & \ {\bf \underline{1}} \
 & \ {\bf \underline{1}} \
\\\hline
 \ $\text{U}(1)_Y$ \
 & \ $-1/2$ \
 & \ $1/6$ \
 & \ $1/2$ \
 & \ $-1$ \
 & \ $2/3$ \
 & \ $-1/3$ \
 & \ $-1/3$ \
 & \ $-2/3$ \
\\\hline\hline
 \ Spin \
 & \ $1/2$ \
 & \ $1/2$ \
 & \ $0$ \
 & \ $1/2$ \
 & \ $1/2$ \
 & \ $1/2$ \
 & \ $0$ \
 & \ $0$ \
\\\hline
 \ $L\#$ \
 & \ $1$ \
 & \ $0$ \
 & \ $0$ \
 & \ $1$ \
 & \ $0$ \
 & \ $0$ \
 & \ $1$ \
 & \ $0$ \
\\\hline
 \ $B\#$ \
 & \ $0$ \
 & \ $1/3$ \
 & \ $0$ \
 & \ $0$ \
 & \ $1/3$ \
 & \ $1/3$ \
 & \ $1/3$ \
 & \ $2/3$ \
\end{tabular}
\end{center}
\caption{
 List of particle contents of the model.
}
\label{tab:particle}
\end{table}

 The particle contents of the colored radiative seesaw model 
are shown in Table~\ref{tab:particle}.
 The model is briefly mentioned in Ref.~\cite{Ref:cZBM}.
 The model includes a scalar leptoquark multiplet~($S_\LQ$)
whose lepton number and baryon number are
$1$ and $1/3$, respectively.
 Under the SM gauge group,
the $S_\LQ$ is assigned to the same representation
of right-handed down-type quarks;
a $\underbar{\bf 3}$ representation of $\SU(3)_C$,
a singlet under $\SU(2)_L$, and hypercharge $Y=-1/3$.
 We also introduce a scalar diquark multiplet~($S_\DQ$)
which has a baryon number $2/3$.
 We take $S_\DQ$ as a $\underbar{\bf 6}$ representation of $\SU(3)_C$,
a singlet under $\SU(2)_L$, and a $Y=-2/3$ field.
 The diquark of a $\underbar{\bf 6}$ representation can be expressed
in a symmetric matrix form as
\begin{align}
S_\DQ
&=
 \begin{pmatrix}
  \red{S_{\DQ 1}}
   &  \ \rg{S_{\DQ 4}}/\sqrt{2} \
   &  \ \rb{S_{\DQ 5}}/\sqrt{2} \ \\
  \ \rg{S_{\DQ 4}}/\sqrt{2} \
   & \green{S_{\DQ 2}}
   &  \ \gb{S_{\DQ 6}}/\sqrt{2} \ \\
  \ \rb{S_{\DQ 5}}/\sqrt{2} \
   & \ \gb{S_{\DQ 6}}/\sqrt{2} \
   & \blue{S_{\DQ 3}}
 \end{pmatrix} .
\label{Eq:6rep}
\end{align}

 The baryon number conservation is imposed to the model
such that the proton decay is forbidden.
 We introduce the soft-breaking term~(see the next paragraph)
of the lepton number conservation to the scalar potential
in order to generate Majorana neutrino masses.
 The Yukawa interactions with the leptoquark and diquark,
which preserve both of the lepton number and the baryon number,
are given by
\begin{align}
{\mathcal L}_\text{Yukawa}
&=
 - \left\{
    \overline{L_\ell^c}\, (Y_L)_{\ell\qd}\, i\,\sigma_2\,Q_\qd^\alpha\,
    + \overline{(\ell_R)^c}\, (Y_R)_{\ell i}\, u_{iR}^\alpha\,
   \right\}
  (S_\LQ^\alpha)^* 
 - \overline{(d_{iR}^{\,\alpha})^c}\,
  (Y_s)_{\qd\qdp}\, d_{jR}^{\,\beta}\,
  (S_\DQ^{\alpha\beta})^*
 + \text{H.c.} ,
 \label{Eq:Lag}
\end{align}
where $\sigma_a$~$(a=1\text{--}3)$ are the Pauli matrices, 
$\alpha$ and $\beta$~($= \red{r}, \green{g}, \blue{b}$)
denote the color indices;
 for example,
$S_\DQ^{\red{rr}}$ corresponds to $\red{S_{\DQ 1}}$ in Eq.~\eqref{Eq:6rep}.
 We choose the diagonal bases of mass matrices
for the charged leptons and down-type quarks.
 Then,
the $\SU(2)_L$ partner of $d_{iL}^{}$ is described as  
$u_{iL}^\prime = (V_\CKM^\dagger)_{ij}^{}\,u_{jL}^{}$,
where $V_\CKM$ is the Cabibbo-Kobayashi-Maskawa~(CKM) matrix
and $u_j^{} = (u_{jR}^{}, u_{jL}^{})^T$
are mass eigenstates of up-type quarks.
 Mass eigenstates $\nu_{iL}^{}$ of neutrinos are given by
$\nu_{iL}^{} = (U_\MNS^\dagger)_{i\ell}^{}\,\nu_{\ell L}^{}$,
where $U_\MNS$ is the Maki-Nakagawa-Sakata~(MNS) matrix.
 The Yukawa matrices ($Y_L$, $Y_R$, and $Y_s$) are $3 \times 3$ matrices under
the lepton flavor~($\ell = e, \mu, \tau$) and the down-type quark flavor~($\qd, \qdp = 1\text{--}3$). 
 While $Y_L$ and $Y_R$ are general complex matrices,
$Y_s$ is a symmetric matrix ($Y_s^T=Y_s$).
 Note that
neutrinos interact with the leptoquark only through $Y_L$,
and we will see later that
$Y_R$ is irrelevant to the neutrino mass at the leading order.

 In the scalar potential of the model,
we introduce the following three-point interaction:
\begin{align}
 \mu\, (S_\LQ^\alpha)^*\, (S_\LQ^\beta)^*\,
   S_\DQ^{\alpha\beta}
 + \text{H.c.}
 \label{Eq:Pot}
\end{align}
 The coupling constant $\mu$ softly breaks
the lepton number conservation by two units
while the baryon number is conserved.
 There is no other possible soft-breaking term
of the lepton number and/or the baryon number.
 We can take the $\mu$ parameter as a real positive value
by using the rephasing of $S_\DQ$.
 Considering
radiative corrections to $m_\LQ^{}$ and $m_\DQ^{}$
via the $\mu$ parameter,
perturbativity requires
$\mu \lesssim \text{min}(m_\LQ^{} , m_\DQ^{})$
as discussed in Ref.~\cite{Nebot:2007bc}
for the Zee-Babu model~(ZBM)~\cite{Ref:Zee-Babu}.

\begin{figure}[t]
\centering
\includegraphics[width=8cm]{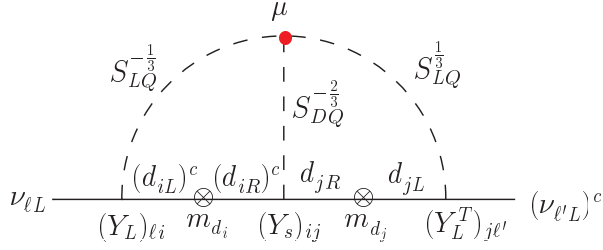}
\caption{
The two-loop diagram for the neutrino mass generation in the model.
}
\label{FIG:mv}
\end{figure}

 The neutrino mass term
$\frac{1}{2} (M_\nu)_{\ell\ell^\prime}\,
\overline{\nu_{\ell L}^{}}\, (\nu_{\ell^\prime L}^{})^c$
in the flavor basis
is generated by a two-loop diagram in FIG.~\ref{FIG:mv}
including the leptoquark and the diquark.
 The mass matrix is calculated as
\begin{align}
(M_\nu)_{\ell\ell'}
=
 + 24 \mu\,
   (Y_L^\ast)_{\ell \qd}^{}\,
   m_{d_i^{}}^{}\,
   (Y_s)_{\qd\qdp}^{}\,
   I_{\qd\qdp}\,
   m_{d_j^{}}^{}\,
   (Y_L^\dag)_{\qdp \ell'}^{},
\label{Eq:mv}
\end{align}
where the loop function $I_{\qd\qdp}$ is defined as
\begin{align}
I_{\qd\qdp}
=
 \int \frac{d^4k_1}{(2\pi)^4}
 \int \frac{d^4k_2}{(2\pi)^4}
  \frac{1}{k_1^2-m_{d_i^{}}^2} \frac{1}{k_1^2-m_\LQ^2}
  \frac{1}{k_2^2-m_{d_j^{}}^2} \frac{1}{k_2^2-m_\LQ^2}
  \frac{1}{(k_1+k_2)^2-m_\DQ^2}.
\end{align}
 The diagram is similar to the one in the ZBM
although $\SU(3)_C$-singlet particles in the loop
are replaced with colored particles.
 Thus,
we refer to this model as the colored Zee-Babu model~(cZBM).
 See, e.g., Refs.~\cite{Nebot:2007bc,Babu:2002uu,Ref:ZBM-lhc}
for studies about the ZBM for comparison with the cZBM\@.
 Note that
$Y_R$ does not contribute to the two-loop diagram.%
\footnote
{
 The $Y_R$ contributes to Majorana neutrino masses
at the higher loop level.
The four-loop contribution
is utilized in the model of Ref.~\cite{Gu:2011ak}
where $Y_L$ is ignored.
}
 In the ZBM,
at least one massless neutrino is predicted
because of the antisymmetric Yukawa coupling matrix. 
 In contrast,
all of three neutrino masses can be non-zero in the cZBM
because $Y_L$ is not an antisymmetric matrix.
 Since new colored scalars
should be much heavier than the SM fermions,
the loop function can be reduced to~\cite{Babu:2002uu}
\begin{align}
I_{\qd\qdp}
\simeq
 I_0
\equiv
 \frac{1}{(4\pi)^4}
 \frac{1}{ (\max[ m_\LQ^{}, m_\DQ^{}] )^2 }
 \frac{\pi^2}{3}
 \tilde{I}( m_\DQ^2/m_\LQ^2),
\end{align}
where 
\begin{align}
\displaystyle \tilde{I}(r)
=
 \begin{cases}
  \displaystyle
  1 + \frac{3}{\pi^2} \{ (\ln r)^2-1 \}
   & \text{for} \quad r \gg 1\\
  1
   & \text{for} \quad r \ll 1
 \end{cases} .
\end{align}

 Hereafter,
we restrict ourselves to the simplest scenario
where $Y_R$ is small enough to be ignored.%
\footnote
{
 If we extend the model as a two-Higgs-doublet model,
we can eliminate the $Y_R$ term
by using a softly-broken $Z_2$ symmetry
(e.g., $u_{iR}^{}$ (or $\ell_R$) and the second Higgs doublet
are $Z_2$-odd fields)
which is also required to avoid
the flavor changing neutral current at the tree level.
 Another example to eliminate the $Y_R$ term is the case
where the leptoquark is not an $\SU(2)_L$-singlet but a triplet.
}
 A benchmark point in the parameter space of the cZBM
is shown in Appendix~\ref{sec:app}.

\section{New colored scalars at the LHC}

\subsection{Leptoquark}

 The main production channel of leptoquarks at hadron colliders
would be the pair-creation from $gg$ and $q\overline{q}$ annihilation%
~\cite{Ref:LQ-pair}.
 The associated production of $S_\LQ$ with a lepton
from $qg$ coannihilation
could also be possible~\cite{Ref:LQ-associate}.
 The pair-production cross section is determined
only by QCD interaction at the leading order~\cite{Ref:LQ-pair}, 
while the associated production mechanism
highly depends on the Yukawa coupling constant of the leptoquark%
~\cite{Ref:LQ-associate}.
 The associated production mechanism is negligible
at a benchmark point shown in the Appendix~\ref{sec:app}
because of tiny $(Y_L)_{\ell 1}$.
 The leptoquarks have been searched at the Tevatron and the LHC\@. 
 The most stringent lower bound on the leptoquark mass
at 95\,\% confidence level
is set as $830\,\GeV$~($840\,\GeV$)
by the recent CMS result at $\sqrt{s}=7\,\TeV$
with $5.0\,\text{fb}^{-1}$ integrated luminosity~\cite{Ref:LQ-cms};
 the pair-production of scalar leptoquarks is assumed
as well as a hundred percent decay branching ratio 
into the first~(second) generation quarks and leptons.
 See also Refs.~\cite{Ref:LQ-1st-atlas,Ref:LQ-2nd-atlas}
for the ATLAS results
with $1.03\,\text{fb}^{-1}$ integrated luminosity.
The analysis of the decay into third generation fermions would be performed in near future. 
The search strategies for the third generation leptoquarks have been studied
in Ref.~\cite{Ref:LQ-3rd}.

 The leptoquark induces various LFV processes. 
 At the tree level,
four-fermion operators
(two left-handed leptons and two left-handed quarks)
are generated by integrating leptoquarks out.
 The constraints on such operators
have been extensively studied in Ref.~\cite{Ref:LQ-lfv-4f}. 
 Tables~3, 4, 12, and 13 in Ref.~\cite{Ref:LQ-lfv-4f}
are relevant to the cZBM\@.
 Especially,
operators
$(\overline{e_L^{}} \gamma^\mu \mu_L^{})(\overline{u_L^{}} \gamma_\mu u_L^{})$
and
$(\overline{\nu_{\ell L}^{}} \gamma^\mu \nu_{\ell^\prime L}^{})
(\overline{d_L^{}} \gamma_\mu s_L^{})$
are strongly constrained by the $\mu$-$e$ conversion search
and the $K$ meson decay measurement,
respectively.
 For the benchmark point given in Appendix~\ref{sec:app},
we have
$|(Y_L)_{e1} (Y_L^\ast)_{\mu 1}|/(4\sqrt{2}\,G_F m_\LQ^2)
= 6.1\times 10^{-11}$
and
$|(Y_L)_{\ell 1} (Y_L^\ast)_{\ell^\prime 2}|/(4\sqrt{2}\,G_F m_\LQ^2)
\lesssim 10^{-7}$
which satisfy constraints shown in Ref.~\cite{Ref:LQ-lfv-4f},
where $G_F = 1.17 \times 10^{-5}\,\GeV^{-2}$.

 At the loop level,
effects of leptoquarks
on charged lepton transitions, i.e., $\ell_i \to \ell_j\gamma$,
have also been studied~\cite{Ref:LQ-lfv-llg}. 
 Since we assume that
$S_\LQ$ has the Yukawa interaction
only with the left-handed quarks (namely $Y_R = 0$),
the contribution from the top quark loop does not give a large enhancement of $\ell_i \to \ell_j\gamma$.%
~\footnote{
 It is known that
the similar process $b\to s \gamma$
(induced  by the uncolored charged Higgs boson)
in the Type-II two-Higgs-doublet model 
is enhanced by the top quark loop~\cite{Ref:bsg}.
}
 Then,
the branching ratio of $\mu \to e \gamma$ is calculated as
\begin{align}
\text{BR}(\mu \to e \gamma)
=
 \frac{ 3 \alpha_\text{EM}^{} }{ 256 \pi G_F^2 m_\LQ^4 }
 \left| \left( Y_L Y_L^\dagger \right)_{e\mu} \right|^2 ,
\label{Eq:meg}
\end{align}
where $\alpha_\text{EM}^{} = 1/137$.
 For example,
a benchmark point shown in Appendix~\ref{sec:app} 
gives $\text{BR}(\mu \to e\gamma) = 6.5\times 10^{-13}$
which satisfies the current upper bound%
~($2.4\times 10^{-12}$ at 90\,\% confidence level)
in the MEG experiment~\cite{Adam:2011ch}.

 Since we take $Y_R = 0$,
the sign of the leptoquark contribution
to the leptonic $g-2$ cannot be changed.
 It is worth to mention that
the contribution of the leptoquark has an appropriate sign
(the plus sign)%
\footnote
{
 For a heavy scalar $\phi$ which interacts
with $\mu_L^{}$ and a light fermion $\psi_L^{}$ as
$\overline{\mu_L^{}} (\psi_L)^c \phi$,
its contribution to the muon $g-2$ has the plus sign
if the electric charge of $\phi$ is greater than $-2/3$.
 See also Ref.~\cite{Kanemitsu:2012dc}.
}
to compensate the difference between
the measured value and the SM prediction for the muon $g-2$.
 The preferred size of $Y_L$ is
$(Y_L Y_L^\dagger)_{\mu\mu} \sim 1$ for $m_\LQ^{} \sim 1\,\TeV$.
 In order to satisfy LFV constraints with this size of $Y_L$,
a simple ansatz is that $Y_L$ is a diagonal matrix.
 Note that
we must take care about the constraint on
$(\overline{\nu_{e L}^{}} \gamma^\mu \nu_{\mu L}^{})
(\overline{d_L^{}} \gamma_\mu s_L^{})$
(see Table~12 in Ref.~\cite{Ref:LQ-lfv-4f})
even if $Y_L$ is diagonal;
 the constraint on $(Y_L)_{e1} (Y_L^\ast)_{\mu 2}$
is difficult to be satisfied with $(Y_s)_{12} \lesssim 1$
because $Y_L$ is related to $Y_s$ through the neutrino mass matrix. 
 We could not find any viable example of such a parameter set
although it might exist with more complicated structures
of $Y_L$ and $Y_s$.

\subsection{Diquark}
 At the LHC,
the diquark $S_\DQ$ in the cZBM
would be singly produced by the annihilation of two down-type quarks.%
\footnote
{
 The diquark can also be created in pair via the gluon-gluon annihilation.
}
 The single production mechanism has an advantage
to search for the relatively heavy diquark due to the $s$-channel resonance. 
 The single production cross section is
determined by $(Y_s)_{11}$, 
which is evaluated in Ref.~\cite{Ref:DQ-prod}
as a function of the diquark mass with a fixed Yukawa coupling constant.
 The $(Y_s)_{11}$ in the cZBM
is less constrained by the neutrino oscillation data
because its contribution to neutrino masses
is suppressed by $m_d^2/m_\DQ^2$.
 If we assume $(Y_s)_{11} = 0.1$ and $m_\DQ^{} = 4\,\TeV$,
the single production cross section $\sigma(dd \to S_\DQ)$
is about $5\,\text{fb}$
at the LHC with $\sqrt{s} = 14\,\TeV$~\cite{Ref:DQ-prod}.
 Note that
the CMS experiment at $\sqrt{s}=7\,\TeV$
with $1\,\text{fb}^{-1}$ integrated luminosity
excludes diquark masses between $1\,\TeV$ and $3.52\,\TeV$
at 95\,\% confidence level
by assuming the diquark decay into two QCD jets
for the $E_6$ diquark
which couples with an up-type quark and a down-type quark%
~\cite{Chatrchyan:2011ns}.
 See also Refs.~\cite{Aaltonen:2008dn,:2012pu,Harris:2011bh}.

 The diquark induces flavor changing neutral current processes 
in the down-type quark sector.
Especially, it gives tree-level contributions to 
$K^0$-$\overline{K^0}$, $B_d^0$-$\overline{B_d^0}$ and $B_s^0$-$\overline{B_s^0}$ mixings,
resulting in strong constraints on $Y_s$.
 By using the notations in Ref.~\cite{Bona:2007vi},
the benchmark point in Eqs.~\eqref{EQ:bench} gives
$\widetilde{C}_K^1 = - (Y_s^\ast)_{11} (Y_s)_{22}/(2 m_\DQ^2) =0$.
 Similarly,
we have $\widetilde{C}_{B_d}^1 = +1.2\times 10^{-12}\,\GeV^{-2}$
and $\widetilde{C}_{B_s}^1 = 0$.
 These values satisfy the constraints obtained in Ref.~\cite{Bona:2007vi}%
~(see also Refs.~\cite{Chakraverty:2000df,Blum:2009sk}).

\begin{figure}[t]
\centering
\includegraphics[width=7cm]{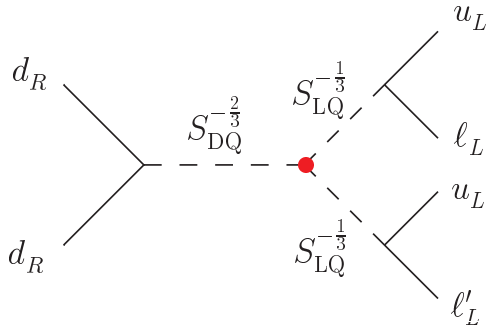}
\caption{The same-signed charged lepton signature without missing energy at the LHC.}
\label{FIG:signature}
\end{figure}

 The diquark in the cZBM
decays into not only a pair of the down-type quarks
but also a pair of leptoquarks.
 The fraction of fermionic and bosonic decay modes 
is calculated as
\begin{align}
\frac{ \sum_{i,j} \Gamma(S_{\DQ 1} \to d_i^{\,r} d_j^{\,r}) }
     { \Gamma(S_{\DQ 1} \to S_\LQ^{\,r} S_\LQ^{\,r}) }
\simeq
 \frac{ m_\DQ^2 \text{tr}(Y_s Y_s^\dagger) }
      { \mu^2 \sqrt{ 1 - \frac{4m_\LQ^2}{m_\DQ^2} } } .
\label{Eq:DQdecay}
\end{align}
 This formula is the same for the other diquarks
because of the $\SU(3)_C$ symmetry.
 We focus on the case where the ratio in Eq.~\eqref{Eq:DQdecay}
is less than about unity
such that the branching ratio for $S_\DQ \to S_\LQ S_\LQ$
becomes ${\mathcal O}(10)\,\%$.
 Subsequently,
$50\,\%$ of each leptoquark decays into
an up-type quark (a down-type quark)
and a charged lepton (a neutrino).
 Then,
the model provides
a characteristic signature in FIG.~\ref{FIG:signature},
whose final state consists of
two QCD jets and the same-signed charged lepton pair
without missing energy.
 The decay chain of the diquark
can be fully reconstructed at the LHC\@.
 This signature can be a smoking gun
for the lepton number violation
because no lepton number is taken away
by invisible particles.
 There is no SM background in principle
because the SM conserves the lepton number.
 It would be also very rare that the SM process mimics the signal process
because the leptons in the signal events are too energetic
to be produced in the SM process.

 It should be emphasized that
the event rate of the process is not necessarily suppressed
though the full process picks up all new coupling constants
relevant to the small neutrino masses
(namely, $Y_s$, $\mu$, and $Y_L$).
 One reason for that is because
the process in FIG.~\ref{FIG:signature} does not have
suppressions with the two-loop factor $1/(16\pi^2)^2$
and with down-type quark masses,
which are used for tiny neutrino masses.
 The other reason is that 
on-shell productions of a diquark and leptoquarks are utilized as
$\sigma(dd \to S_\DQ)
\text{BR}(S_\DQ \to S_\LQ S_\LQ)
\left[\sum_{\ell, i} \text{BR}(S_\LQ \to \ell_L^{} u_{iL}^{})\right]^2$;
 even if a partial decay width
is controlled by a small coupling constant
(e.g., $S_\LQ \to \ell_L^{} u_{iL}^{}$ via $(Y_L)_{\ell i}$),
its branching ratio becomes sizable
when the total decay width is also controlled by small coupling constants.
 In this scenario,
the cZBM seems the new physics model
which is the most easily probed at the LHC
and takes us to the top of the energy frontier.

 For the benchmark point shown in the Appendix~\ref{sec:app},
the ratio in Eq.~\eqref{Eq:DQdecay} is 0.18 
for which 85\,\% of $S_\DQ$ decays into $S_\LQ S_\LQ$.
 Then,
15\,\%~(7\,\%) of $S_\LQ$ decays into
a charm quark~(a top quark)
associated with an electron or a muon.
 Decays into an up quark are negligible
for the bench mark point.
 The decay into a tau lepton
might not be reliable because it gives missing neutrinos.
 Even if $S_\LQ$ decays into a top quark,
hadronic decays~(68\,\%) of $W^\pm$ from the top quark decay
have no missing energy.
 As a result,
the cross section for L\#V events without missing energy
is about $0.18\,\text{fb}$ at the LHC with $\sqrt{s}=14\,\TeV$
for the benchmark point.

 In the energy scale which is much below the new scalar masses,
the diagram in FIG.~\ref{FIG:signature}
becomes a dimension-9 operator of six fermions.
 Such L\#V operators up to dimension-11
have been studied in Refs.~\cite{Ref:cZBM,de Gouvea:2007xp}.
 The dimension-9 operator in the cZBM is highly suppressed 
by the inverse power of mass scales of new colored particles
as well as by new Yukawa coupling constants.
 Therefore
the collider signature in FIG.~\ref{FIG:signature}
does not conflict with the stringent constraints
from the other lepton number violating observable
such as the neutrinoless double beta decay~\cite{Ref:0v2beta}, and lepton number violating rare decays 
$\tau^\pm \to \ell^\mp M^\pm M^\pm (M=\pi, K)$~\cite{Ref:lnv-tau}, 
$M^\pm \to {M'}^\mp \ell^\pm {\ell'}^\pm (M=B, K, D)$~\cite{Ref:lnv-meson}
and $t\to b\ell^+{\ell'}^+W^-$~\cite{Ref:lnv-top,Quintero:2011yh}.

\section{Conclusions}

 We have studied a model for neutrino mass generation
with the scalar leptoquark $S_\LQ$ and the scalar diquark $S_\DQ$. 
 Tiny Majorana neutrino masses are induced at the two-loop level
where the colored particles are involved in the loop.
 The trilinear scalar coupling constant
between two leptoquarks and a diquark
is the only parameter of the lepton number violation in this model. 
 The diquark can be singly produced at the LHC via the resonance mechanism,
and it can decay into a pair of leptoquarks
through the lepton number violating coupling.
 The leptoquarks can further decay into a charged lepton and an up-type quark. 
 Thus,
the model gives a distinctive signature at the LHC,
namely
$pp \to S_\DQ \to S_\LQ S_\LQ \to \ell^- \ell^{\prime -} j j$
without missing energy,
which would be a clear evidence of the lepton number violation.
 We have shown that
the lepton number violating process is not suppressed
because of on-shell productions and decays of the diquark and the leptoquarks,
while Majorana neutrino masses are highly suppressed.

\acknowledgments
 We thank Joe Sato, Eibun Senaha, and Hiroshi Yokoya for valuable comments.
 M.K.\ is supported by the NTU Grant No.~101R7701 and the Laurel program.
 The work of H.S.\ was supported in part by the Grant-in-Aid
for Young Scientists~(B) No.~23740210.
 K.T.\ was supported, in part, by the Grant-in-Aid for Scientific research
from the Ministry of Education, Science, Sports, and Culture~(MEXT), Japan,
No.~23104011. 

\appendix
\section{A benchmark point}\label{sec:app}
 Here,
we show a benchmark point of the model:
\begin{subequations}
\begin{align}
&
Y_L
=
 \left(
 \begin{array}{rrr}
  8.1\times 10^{-5}
   & 4.0\times 10^{-2}
   & - 7.0\times 10^{-3}\\
  -4.9\times 10^{-5}
   & 5.3\times 10^{-2}
   & 4.4\times 10^{-2}\\
 3.1\times 10^{-5}
  & -2.3\times 10^{-2}
  & 8.9\times 10^{-2}
 \end{array}
 \right) ,
\label{EQ:YL-bnech}\\
%
&
Y_s
=
 \begin{pmatrix}
  1.0\times 10^{-1}
   & 0
   & 0\\
  0
   & 0
   & -1.2\times 10^{-2}\\
 0
  & -1.2\times 10^{-2}
  & -3.8\times 10^{-4}
 \end{pmatrix} ,
\label{EQ:Ys-bench}\\
%
&
 \mu = 1\,\TeV , \quad
 m_\LQ^{} = 1\,\TeV , \quad
 m_\DQ^{} = 4\,\TeV .
\end{align}\label{EQ:bench}
\end{subequations}
 We define an overall constant of neutrino masses
as $C \equiv 24 \mu I_0 \simeq 6.0\times 10^{-7}\,\GeV^{-1}$.

 The neutrino mass matrix is diagonalized as
$U_\MNS^\dagger M_\nu U_\MNS^\ast
= \text{diag}(m_1, m_2 e^{i\alpha_{21}^{}}, m_3 e^{i\alpha_{31}^{}})$
with $U_\MNS$ which can be parametrized as
\begin{align}
&
U_\MNS
=
\begin{pmatrix}
 1 & 0 & 0\\
 0 & c_{23} & s_{23}\\
 0 & -s_{23} & c_{23}
\end{pmatrix}
\begin{pmatrix}
 c_{13} & 0 & s_{13} e^{-i\delta}\\
 0 & 1 & 0\\
 -s_{13} e^{i\delta} & 0 & c_{13}
\end{pmatrix}
\begin{pmatrix}
 c_{12} & s_{12} & 0\\
 -s_{12} & c_{12} & 0\\
 0 & 0 & 1
\end{pmatrix} ,
\end{align}
where $c_{ij}$~($s_{ij}$) denotes $\cos\theta_{ij}$~($\sin\theta_{ij}$).
 We use the following values:
$\sin^2{2\theta_{23}} = 1$,
$\sin^2{2\theta_{13}} = 0.1$,
$\sin^2{2\theta_{12}} = 0.87$,
$\delta = 0$,
$\Delta m^2_{31} = 2.4\times 10^{-3}\,\text{eV}^2 >0$,
and $\Delta m^2_{21} = 7.6\times 10^{-5}\,\text{eV}^2$.
 Matrices $Y_L$ and $Y_s$ in Eqs.~\eqref{EQ:bench}
are constructed by assuming the following structures:
\begin{subequations}
\begin{align}
&
Y_L
=
U_\MNS^\ast
\begin{pmatrix}
 1
  & 0
  & 0 \\[2mm]
 0
  & \displaystyle \sqrt{\frac{m_3}{m_2+m_3}}
  & -\displaystyle \sqrt{\frac{m_2}{m_2+m_3}}\\[5mm]
 0
  & \displaystyle \sqrt{\frac{m_2}{m_2+m_3}}
  & \displaystyle \sqrt{\frac{m_3}{m_2+m_3}}
\end{pmatrix}
\begin{pmatrix}
 x & 0 & 0\\
 0 & y & 0\\
 0 & 0 & z
\end{pmatrix} , \\
&
Y_s
=
\begin{pmatrix}
 (Y_s)_{11}
  & 0
  & 0\\[2mm]
 0
  & 0
  & \displaystyle -\frac{\sqrt{m_2 m_3}}{y z m_s m_b C}\\[4mm]
 0
  & \displaystyle -\frac{\sqrt{m_2 m_3}}{y z m_s m_b C}
  & \displaystyle -\frac{m_3 - m_2}{ z^2 m_b^2 C}
\end{pmatrix} .
\end{align}
\end{subequations}
 It is easy to see that
$M_\nu$ with these matrices results in
$m_1 = x^2 m_d^2 C (Y_s)_{11}$,
$\alpha_{21}^{}=0$, and $\alpha_{31}=\pi$.
 We use
$x=10^{-4}$, $y=0.07$, $z=0.1$,
$m_d=5\times 10^{-3}\,\GeV$, $m_s=0.1\,\GeV$, $m_b=4.2\,\GeV$
and $(Y_s)_{11} = 0.1$.
 Note that the benchmark point gives
$(M_\nu)_{ee} \simeq 1.5\times 10^{-3}\,\text{eV}$,
which is the effective mass relevant for the neutrinoless double beta decay.


\end{document}